\documentclass[10pt,twocolumn]{article}

\usepackage[letterpaper,top=0.75in,bottom=1in,left=0.625in,right=0.625in,columnsep=0.25in]{geometry}
\usepackage{mathptmx}
\usepackage[T1]{fontenc}
\usepackage[utf8]{inputenc}
\makeatletter
\renewcommand\section{\@startsection{section}{1}{\z@}{-2.2ex plus -1ex minus -.2ex}{1.2ex plus .2ex}{\normalfont\normalsize\scshape\centering}}
\renewcommand\subsection{\@startsection{subsection}{2}{\z@}{-1.8ex plus -1ex minus -.2ex}{0.8ex plus .2ex}{\normalfont\normalsize\itshape}}
\renewcommand\subsubsection{\@startsection{subsubsection}{3}{\z@}{-1.5ex plus -1ex minus -.2ex}{0.5ex plus .2ex}{\normalfont\normalsize\itshape}}

\makeatother
\AtBeginDocument{\setlength{\abovedisplayskip}{5pt}\setlength{\belowdisplayskip}{5pt}\setlength{\abovedisplayshortskip}{3pt}\setlength{\belowdisplayshortskip}{3pt}}
\usepackage{amsmath,amssymb}
\usepackage{booktabs}
\usepackage{graphicx}
\usepackage{microtype}
\usepackage{xcolor}
\usepackage{tikz}
\usetikzlibrary{arrows.meta,positioning,fit,backgrounds,calc}
\usepackage{algorithm}
\usepackage{algorithmic}
\usepackage[font=small,labelfont=sc]{caption}
\usepackage{url}
\usepackage[hidelinks]{hyperref}
\usepackage{balance}

\newcommand{\pplus}{p^{+}}
\newcommand{\gminus}{g^{-}}
\newcommand{\vplus}{v^{+}}

\title{\LARGE\bfseries AI-Driven, Timing-Aware Power Optimization:\\[2pt]
\large Learned Ranking of Post-Route Power Recovery and Baseline-Preserving Repair}
\newif\ifanon \anonfalse
\ifanon
  \author{Anonymous Author(s)\\ \normalsize Affiliation withheld for review}
\else
  \author{Amit Sharma\\ \normalsize IEEE Member\\ \normalsize \href{mailto:amit.sharma.ai@ieee.org}{amit.sharma.ai@ieee.org}}
\fi
\date{}

\begin{document}
\maketitle
\thispagestyle{empty}

\begin{abstract}
Post-route power recovery changes cell choices under tightly coupled timing, electrical and physical constraints, and the value of each change depends on what it costs to repair its side effects. We propose an AI-driven framework in which a learned model drives both phases of the optimization: it ranks the power-recovery edits to try and the repair actions that follow, while deterministic electronic design automation (EDA) analyses govern every acceptance. The framework records baseline constraint margins, identifies the damage a recovery transaction introduces in each analysis view, holds pre-existing violations at their baseline rather than working them inside the power step, and caps repair spending at a fraction of the measured saving. An ensemble of gradient-boosted models predicts power change, normalized damage reduction and constraint-regression risk from action and timing-context features available before each trial; calibrated, conservative estimates feed separate acquisition scores for recovery and repair, with a deterministic fallback when the model is uncertain. Recovery and repair remain speculative until physical implementation and fresh analysis establish baseline preservation and positive net saving. We specify the contract, the learned policy, the transactional flow and a pre-registered public evaluation protocol on open designs. Evidence from a companion study on public designs supports the contract; the learned policy is proposed, and no trained-model results or proprietary design measurements are reported.
\end{abstract}


\section{Introduction}
Cell sizing and threshold-voltage (Vt) assignment are among the last ways to reduce power in physical design, when the scope for structural change is small and every edit must survive signoff. A lower-leakage or smaller cell reduces power but also changes delay, input loading, output slew and the timing requirements of sequential cells it drives. Repairing those side effects can consume the saving. Established recovery and sizing methods manage the interaction through sensitivity calculations, incremental timing updates and constrained cell changes \cite{fatemi2019,kahng2013,hu2012,flach2014}.

The repair objective matters most when the implementation enters the power step with violations of its own. An ordinary closure command, aimed at every violator with a target slack of zero, repairs those pre-existing violations together with the damage the recovery created. Its power cost then combines the design's general timing closure with the cost of retaining the new saving, and the two are indistinguishable in the result. We therefore separate \emph{baseline preservation}, returning every constraint to the margin it had before the recovery, from the stronger requirement that the whole design become timing-clean. The separation is an accounting and constraint policy, not permission to weaken final implementation requirements. A companion study on public designs, summarized in Section~\ref{sec:evidence}, found that the retention of a conventional whole-design repair fell steeply as the ratio of pre-existing violators to recovery damage rose, while the same command aimed only at the damage held nearly all of the saving \cite{companion}. That evidence motivates making the contract explicit; it says nothing yet about learning.

Learning enters because both phases are sequential decisions over a large discrete action space in which every evaluation is an expensive EDA call. Whether a swap saves power without damage, and whether a repair restores a margin cheaply, depends on circuit context: slews, loads, path sharing and the sequential requirements of neighbouring cells. The order in which candidates are trialed therefore decides how much verification work is spent per unit of retained saving. The companion study found the order in which recovery edits were restored to be a first-order determinant of give-back at equal residual damage; that sensitivity is the opening for a learned policy that estimates, before a trial, how much power an edit will save, how much damage it will cause or remove, and how likely it is to regress a protected check. \emph{AI-driven} here means that the model decides what to try, in which order and when to stop. It does not mean that the model certifies the result: the same deterministic candidate generator, constraints and final verification apply to learned and heuristic policies alike, which makes the contribution of learning experimentally separable from the contribution of damage scoping or a better repair recipe.

The paper makes three contributions. First, a baseline-relative optimization contract that couples recovery and repair through per-view damage sets, baseline targets, a budget tied to the measured saving, and complete transaction rollback (Section~\ref{sec:contract}). Second, a learned, uncertainty-aware ranking policy that drives both recovery and repair, with calibrated conservative estimates, a deterministic fallback and a charged adaptation mode (Section~\ref{sec:policy}). Third, a transactional flow (Section~\ref{sec:flow}) and a pre-registered public evaluation protocol (Section~\ref{sec:eval}) with hypotheses, arms, metrics and artifacts that separate the value of learning from the value of the contract. The architecture and protocol are proposed; the evidence of Section~\ref{sec:evidence} is cited from the companion study, and no empirical claim from any deterministic flow is attributed to the learned policy.

\section{Background and Related Work}
\subsection{Recovery and repair in practice}
Sensitivity-based industrial recovery weighs power against timing, multi-corner behaviour and electrical constraints \cite{fatemi2019}. Signoff-timer sizing \cite{kahng2013,hu2012}, Lagrangian-relaxation sizing with Vt assignment \cite{flach2014,ozdal2011} and incremental sizing with an ECO-cost objective \cite{leegupta2013} form the sizing core on which recovery flows rest. Commercial signoff tools offer power-aware ECO fixers and recovery commands: one guarantees that existing violations do not worsen and backs out changes that create new ones \cite{ptdocs}; another documents a swap-only setup recovery after a large power optimization \cite{tempusdocs}. Re-speeding only the cells changed during recovery, one Vt level at a time, appears in a signoff-timer utility and a public patent application \cite{zahn2008,zahn2010}. The open-source resizer guards the design's worst slack and can undo its own recovery moves through a journal \cite{openroad}. Selective restoration and incremental sizing are therefore building blocks, not novelty claims, in this proposal.

\subsection{Learned sizing and learned ECO}
Graph learning has been applied to signoff power optimization through Vt assignment \cite{lu2020} and to discrete gate sizing on directed VLSI graphs \cite{dagsizer2023}; transformer-based \cite{transsizer2022} and LLM-enhanced differentiable \cite{legosize2025} sizers, physically aware large-scale sizing \cite{ye2025} and disentangled differentiable timing--power co-optimization \cite{pan2026} extend the family, as do fast differentiable timers \cite{insta2025} and reinforcement learning for IR-aware ECO timing \cite{jiang2024}. These works motivate learning from circuit context and serve as baselines where action spaces are compatible; they do not establish the effectiveness of the architecture proposed here. The proposed distinction is empirical, uncertainty-aware ranking of recovery and repair actions under an explicit retained-saving budget and a baseline-relative constraint contract, with deterministic acceptance. Gradient-boosted trees \cite{xgboost} are a practical first model family for mixed numerical and categorical features; the choice is an implementation hypothesis to test, not a claim of superiority over graph or transformer models.

\subsection{Companion deterministic study}
A companion manuscript \cite{companion} states a damage-aimed repair policy and applies it with signoff fixers, with an unmodified open-source resizer steered through path exceptions, and with restoration of the recovery's own edits, on public designs and on proprietary blocks. The present paper generalizes its contract to arbitrary constraint types with declared tolerances, replaces its fixed rankings with a learned, uncertainty-aware policy that also drives the recovery phase, and defines the public protocol under which the value of learning can be measured. Only the companion study's public-design evidence is used here.

\subsection{Scope}
The initial implementation targets public infrastructure \cite{openroad,opensta,asap7}; its action space is functionally equivalent cell substitution and localized repair, including buffer insertion where physical data supports it. Clock architecture, logic restructuring and functional-constraint changes are out of scope. A public predictive technology supports reproducible research but not foundry-qualified signoff \cite{asap7}.

\section{Optimization Contract}\label{sec:contract}
\subsection{States, margins and baseline targets}
Let $x_0$ be an accepted implementation checkpoint, $x_1$ the speculative recovery-only state and $x_2$ the repaired candidate offered for commitment. All power measurements used in one comparison share the activity model, analysis scenario, library interpretation, power components and measurement fidelity; denote the powers by $P_0$, $P_1$ and $P_2$.

Index each constrained item by $j$, which identifies its endpoint or pin, analysis view and check type, and let $m_j(x)$ be its signed margin, positive when the constraint is satisfied. For setup and hold this is slack; for a maximum-transition or maximum-capacitance constraint it is the applicable limit minus the measured quantity. Different check types keep their own physical units. The baseline target and the normalized damage are
\begin{equation}
t_j = \min\{0,\, m_j(x_0)\}, \qquad
d_j(x) = \frac{\max\{0,\; t_j - m_j(x) - \tau_j\}}{a_j},
\label{eq:damage}
\end{equation}
where $\tau_j \ge 0$ is a declared numerical tolerance and $a_j > 0$ a fixed scale in the units of the margin; both are chosen before testing and held constant across compared methods. A constrained item created by the repair itself (a new buffer's pins, for instance) has target zero and cannot inherit a waiver from an unrelated baseline pin. Technology-specific hard legality requirements are never relaxed by the baseline rule.

The current damage set and the aggregate search objective are
\begin{equation}
D(x) = \{\, j : d_j(x) > 0 \,\}, \qquad
\Phi(x) = \sum_j w_j\, d_j(x), \quad w_j > 0 .
\label{eq:phi}
\end{equation}
The weights and scales prevent unlike units from being summed directly; the aggregate guides the search, while acceptance checks every item separately, so a better total negative slack cannot conceal a newly damaged endpoint. Baseline-relative feasibility requires $m_j(x_2) \ge t_j - \tau_j$ for every required check. A separate timing-clean experiment uses zero targets throughout and reports its additional power cost. Across successive transactions, the run's original baseline targets are enforced at every commit, so tolerances cannot accumulate into baseline drift; per-round power accounting may use the current accepted checkpoint.

\subsection{Power accounting and repair budget}
Define the provisional recovery saving $S = P_0 - P_1$. A budget and a retention ratio are meaningful only when $S > 0$:
\begin{equation}
B = \beta S, \quad 0 \le \beta < 1, \qquad
C = P_2 - P_1, \qquad
R = \frac{P_0 - P_2}{S}.
\label{eq:budget}
\end{equation}
If the measured give-back satisfies $C \le B$, then $R \ge 1 - \beta$. This is an algebraic accounting identity under consistent measurements; it neither proves that a feasible repair exists nor certifies any prediction. If $S \le 0$ the round is rejected or handled as ordinary timing optimization, and no retention value is reported.

Each repair trial is charged by the measured total-power change from the current checkpoint, and cumulative spend is re-evaluated against $x_1$ rather than reconstructed from historical per-cell savings: slew, load, activity propagation and physical changes make those estimates non-additive. A configurable minimum net saving separates a meaningful improvement from numerical variation.

\subsection{Why the target population sets the cost}
With target zero on every violator, the repair's spend decomposes into the cost of restoring the recovery's damage and the cost of closing violations that pre-dated it; the second term scales with the pre-existing population and is paid in power whatever the repair move. The contract removes it by construction: pre-existing violators are held at $t_j$ and cost nothing unless the recovery worsened them. The companion study's public-design evidence for this decomposition is summarized in Section~\ref{sec:evidence}.

\subsection{Feasibility and search boundaries}
Recovery may create bounded temporary damage inside an isolated trial branch with an explicit damage cap, edit limit and analysis-call limit selected on validation designs. These limits restrict search effort, not timing or physical requirements, and a damaged state is never installed as the accepted implementation. At trial start a protected check satisfies $m_j(x) \ge t_j - \tau_j$; regression means crossing below that threshold during the trial. During repair, protected checks must stay within their targets, and an accepted action must reduce normalized damage or improve power without increasing damage. All relevant checks are refreshed after accepted changes, including checks outside the initial damage set. If no acceptable action remains, the branch is discarded. Constraint exceptions used to steer an optimizer are removed before independent verification under the original constraints.

\section{Learned Policy}\label{sec:policy}
\subsection{Decision problem}
Each phase is a sequential selection over a legal action set whose evaluation requires an incremental timing and power update, and whose outcome depends on the state the previous actions left behind. We frame the policy as learned ranking with uncertainty rather than end-to-end reinforcement learning: every step is verified, so per-step labels are free; complete trajectories per design are few, which favours sample-efficient supervised surrogates; and a ranking over enumerated legal actions is auditable. A reinforcement-learning policy with a compatible action space is retained as a comparison arm (Section~\ref{sec:eval}).

\subsection{Candidate generation and features}
A deterministic generator enumerates legal actions before learning is applied. Recovery actions are permitted lower-leakage substitutions and drive reductions; repair actions are restoration of a recorded recovery edit, permitted drive or Vt changes, and localized buffer insertion when physical data supports it. The generator enforces functional equivalence, site and footprint compatibility, library exclusions and user-declared edit restrictions. Candidate features (Table~\ref{tab:features}) describe the current state and the proposed action before a trial and exclude the outcome being predicted. They use generic functional and electrical attributes rather than design identifiers or proprietary cell names. Features are refreshed whenever the implementation changes; stale path incidence and slack values are never reused as if they described the new state, and feature-extraction cost is charged to the reported runtime.

\begin{table}[t]
\centering\footnotesize
\caption{Proposed input features. No trained feature set or measured predictive accuracy is reported.}
\label{tab:features}
\begin{tabular}{@{}p{0.3\columnwidth}p{0.62\columnwidth}@{}}
\toprule
Feature group & Examples available before the trial \\
\midrule
Action and cell & Action type, equivalent-cell class, relative drive, area, characterized leakage, Vt class delta \\
Electrical context & Input slew, output load, fanout, activity, local interconnect estimate \\
Timing context & Setup and hold margins by view, baseline targets, affected-path incidence, shared-cell count \\
Recovery history & Whether the edit restores a prior action; its recorded measurement context and trial cost \\
Search state & Current damage $\Phi$, remaining budget, candidate count, analysis-call allowance \\
\bottomrule
\end{tabular}
\end{table}

\subsection{Trial labels and model fitting}
For candidate $a$ at state $x$, let $x' = x \oplus a$ be the EDA-evaluated trial. Its learning targets are
\begin{equation}
\begin{aligned}
y_P &= \frac{P(x') - P(x)}{P_0}, \qquad y_G = \Phi(x) - \Phi(x'),\\
y_V &= \mathbf{1}\{\text{a protected check regresses in } x'\}.
\end{aligned}
\label{eq:labels}
\end{equation}
The power and damage regressors use the Huber loss; the regression-risk classifier uses binary cross-entropy with calibration assessed separately. Several bootstrap-trained replicas of each model supply a disagreement measure. Training records include accepted edits, rejected trials and rollbacks, since successful edits alone would hide the examples most relevant to constraint risk; the protected set is computed from the pre-action state $x$; tool failures carry a failure flag, and unavailable labels are masked rather than treated as successful outcomes. Incremental-analysis labels and physically implemented labels carry distinct fidelity tags and are not pooled without testing the resulting bias. The primary dataset is generated from public designs by reproducible scripted trials. All derived netlists, layout seeds, corners, snapshots and trajectories of one design family stay in one partition; training, model selection, uncertainty calibration and final testing use disjoint design groups, and normalization and feature selection are fitted on training data only.

\subsection{Uncertainty-aware acquisition}
Let $\pplus$ be an upper estimate of normalized power change, $\gminus$ a lower estimate of damage reduction and $\vplus$ a conservative estimate of regression risk. Residual margins selected on held-out calibration designs adjust $\pplus$ and $\gminus$; the classifier is calibrated separately and $\vplus$ is clipped to $[0,1]$. These are ranking aids, not probabilistic guarantees for unseen technologies or adaptively selected actions. Recovery and repair use different acquisition scores:
\begin{align}
A_{\mathrm{rec}}(a) &= -\pplus - \lambda\,\max(0, -\gminus) - \eta\,\vplus, \label{eq:arec}\\
A_{\mathrm{rep}}(a) &= \frac{\max(0, \gminus)}{\varepsilon + \max(0, \pplus)} - \eta\,\vplus . \label{eq:arep}
\end{align}
Recovery favours the largest conservative power reduction with limited predicted damage; repair favours deficit reduction per unit of predicted give-back. The constant $\varepsilon > 0$ stabilizes near-zero predicted costs, and the non-negative weights $\lambda$, $\eta$ are chosen on validation designs and then frozen. Power-reducing repair actions with equal damage reduction may be retained as a secondary choice under the same measured checks. After every accepted edit the features are refreshed and the candidates re-ranked. Recovery stops when the conservative expected saving of the best remaining candidate falls below a declared fraction of the analysis cost per trial, or at the edit or call limit; repair stops when the damage set is empty or the budget would be exceeded.

\subsection{Fallback and cold start}
High ensemble disagreement, feature values outside the calibrated range or failed calibration diagnostics trigger a deterministic fallback. For repair the fallback ranks legal restoration actions by measured deficit coverage per unit of estimated give-back, re-read after each batch, which is the companion study's incidence ranking; for recovery it is the sensitivity order of the underlying engine. Fallback decisions are logged separately from model decisions. A cold-start run without a trained model is reported as a heuristic run and is not evidence of learned optimization.

\subsection{Adaptation policy}
The primary evaluation freezes the trained model on test designs. A separate online-adaptation experiment may update the model from trial labels observed earlier in the same trajectory, charging those updates to runtime, never accessing future outcomes and preserving the held-out split. No exploratory action bypasses the deterministic acceptance checks.

\subsection{Place in agentic flows}
The transaction of Section~\ref{sec:flow} is exposed as a tool with a declared contract: inputs, limits, a budget share and a verification report. An orchestrating agent, including one based on a large language model, may schedule transactions, choose budget shares within policy and summarize damage by apparent cause for human review; it does not select cells, set targets outside the contract or certify timing.

\section{Transactional Optimization Flow}\label{sec:flow}
The learned model controls candidate priority; EDA tools supply the measurements that control state transitions (Fig.~\ref{fig:arch}). The separation matters because low prediction error does not establish that a sequence of edits is feasible, and every committed round must meet the same acceptance contract whichever policy proposed it.

\begin{figure*}[t]
\centering
\begin{tikzpicture}[
  font=\footnotesize,
  box/.style={draw, rounded corners=2pt, align=center, minimum height=8mm, inner sep=3pt, text width=26mm, fill=white},
  ai/.style={box, fill=blue!8, draw=blue!50!black},
  ok/.style={box, fill=green!8, draw=green!40!black},
  bad/.style={box, fill=red!6, draw=red!50!black},
  arr/.style={-{Latex[length=2mm]}, thick},
  dasharr/.style={-{Latex[length=2mm]}, thick, dashed, gray},
  lab/.style={font=\scriptsize, text=black!70}
]
\node[box] (base) {\textbf{Baseline state}\\ constraints, margins, $P_0$};
\node[box, right=7mm of base] (gen) {\textbf{Candidate generation}\\ legal cell-level edits};
\node[ai, right=7mm of gen] (rank) {\textbf{Learned ranking}\\ ensemble, $A_{\mathrm{rec}}$ / $A_{\mathrm{rep}}$};
\node[box, right=7mm of rank] (trial) {\textbf{EDA trial}\\ incremental timing and power};
\node[ai, above=5mm of rank] (train) {\textbf{Training / replay}\\ offline; frozen at test};
\node[ai, below=8mm of rank] (repair) {\textbf{Scoped repair}\\ AI-ranked trials in $D(x_1)$, iterated per batch};
\node[bad, below=8mm of trial] (roll) {\textbf{Rollback}\\ restore checkpoint};
\node[box, below=8mm of gen] (phys) {\textbf{Physical execution}\\ legalize, route, extract, re-time};
\node[ok, below=8mm of base] (commit) {\textbf{Verified checkpoint}\\ baseline kept, $C \le B$, $R>0$};
\draw[arr] (base) -- (gen);
\draw[arr] (gen) -- (rank);
\draw[arr] (rank) -- (trial);
\draw[arr] (trial.south) -- ++(0,-3mm) -| (repair.north) node[lab, pos=0.22, below] {trial passes initial checks};
\draw[arr] (trial) -- (roll) node[lab, midway, right] {trial fails};
\draw[arr] (repair) -- (phys);
\draw[arr] (phys) -- (commit);
\draw[arr] (repair.east) -- ++(4mm,0) |- (roll.west);
\draw[dasharr] (trial.north) |- (train.east) node[lab, pos=0.75, above] {trial outcomes};
\draw[dasharr] (train) -- (rank);
\draw[arr] (phys.south) -- ++(0,-4mm) -| (roll.south) node[lab, pos=0.3, below] {final validation fails $\rightarrow$ rollback};
\begin{scope}[on background layer]
\node[draw=black!30, rounded corners=3pt, fill=black!2, fit=(gen)(trial)(repair)(roll), inner sep=5pt, label={[lab]above left:private trial branch}] {};
\end{scope}
\end{tikzpicture}
\caption{Proposed architecture: AI-ranked optimization with deterministic acceptance. Recovery and repair run inside a trial branch; only a physically verified result that preserves every baseline target within the repair budget becomes a new accepted checkpoint. Shaded blue blocks are the learned components; everything else is deterministic.}
\label{fig:arch}
\end{figure*}

\begin{algorithm}[t]
\caption{Proposed optimization transaction}
\label{alg:txn}
\begin{algorithmic}[1]
\STATE Load a verified checkpoint and the original constraints; record baseline margins $m_j(x_0)$, consistent power $P_0$ and hard legality requirements.
\STATE Generate legal recovery candidates; rank them by $A_{\mathrm{rec}}$, or by the declared fallback when the model is unavailable or outside its calibrated domain.
\STATE Trial candidates one at a time in a saved branch; refresh timing and power; keep only changes within the provisional search limits and roll back the rest. After each accepted edit, refresh features and re-rank. Stop at the edit or call limit or when no acceptable candidate remains.
\STATE Freeze the recovery-only state $x_1$. If $S \le 0$, discard the round. Otherwise set $B = \beta S$ and compute $D(x_1)$ against the baseline targets.
\STATE If $D(x_1)$ is empty, go to step 7. Otherwise generate restoration and repair candidates, rank them by $A_{\mathrm{rep}}$, trial each, measure cumulative spend, and check progress, protected constraints and legality; keep an acceptable action, else restore the preceding state and try the next.
\STATE Refresh the damage set and features after every accepted action. If damage remains when the budget, progress or call limits are reached, discard the whole round and return the original checkpoint.
\STATE Implement the surviving edit set, including repair buffers: legalize, route as required, extract parasitics, verify functional equivalence, and run fresh timing, electrical, physical and power analyses under the original constraints.
\STATE Commit only if every baseline target and hard requirement passes and the net saving exceeds the declared minimum; check the final budget at comparable fidelity. Otherwise roll back the whole round.
\end{algorithmic}
\end{algorithm}

Algorithm~\ref{alg:txn} uses single-action trials so that charging and rollback are unambiguous; all trials, failures and analysis costs are logged. Batching is an optional optimization whose invalid batches must be replayed or subdivided to a valid prefix, with the extra work charged. The change journal must cover inserted cells, nets, constraints and physical state; restoring only the cell master list is insufficient after buffer insertion.

Final retention requires a comparable recovery-only reference; if that reference cannot be implemented and evaluated consistently, the budgeted experiment records the budget as unevaluable, and a separately labelled net-saving-only variant may report final power reduction without claiming a verified budget. A failed transaction returns an explicit no-improvement result, never a partially repaired deliverable.

\section{Evidence for the Contract on Public Designs}\label{sec:evidence}
The companion study \cite{companion} tested the contract's central claim without any learned component, on four OpenROAD example designs (\textit{aes}, \textit{ibex}, \textit{jpeg}, \textit{ethmac}) placed and routed on the ASAP7 predictive technology \cite{openroad,asap7}, 13{,}313 to 58{,}035 instances. A setup uncertainty on every clock raised the pre-existing violator population while the recovery was held fixed, so the population was the only variable; the stock whole-design repair command was compared with the identical command aimed at the damage set through path exceptions (false paths on untouched pre-existing violators, maximum-delay targets on damaged endpoints), removed before verification. Table~\ref{tab:evidence} summarizes the result. Retention of the stock repair fell with the ratio of pre-existing violators to damage (Spearman $-0.74$, $p = 0.006$, over twelve re-run levels; within-series mean $-0.85$ over forty fixed-recovery levels), while the aimed arm was flat ($-0.11$), left no damage residual at any level and was ahead of the stock repair by more than $0.02$ at ten of twelve levels and within $0.02$ at the other two. On the hardest \textit{aes} level the stock repair spent roughly 110 times the saving chasing 384 pre-existing violators it never closed.

\begin{table}[t]
\centering\footnotesize
\caption{Companion-study evidence on public designs \cite{companion}: the same open-source resizer command, aimed at every violator (stock) or at the damage set (aimed). \textit{aes} on ASAP7; ratio = pre-existing setup violators $\div$ setup endpoints the recovery damaged. No learned policy was involved.}
\label{tab:evidence}
\begin{tabular}{@{}lccc@{}}
\toprule
\textit{aes} constraint level & ratio & stock repair $R$ & aimed repair $R$ \\
\midrule
Level 1 & 4.5 & 0.44 & 0.95 \\
Level 2 & between & $-17.5$ & 0.90 \\
Level 3 & 24 & $-108.9$ & 0.63 \\
\midrule
\multicolumn{4}{@{}p{0.95\columnwidth}@{}}{\footnotesize Four designs, three levels each: aimed ahead at 10 of 12 levels, within 0.02 at 2. Caveats: at 6 of the 12 re-run levels the router failed inside the stock repair, which finished on placement parasitics (likely understating its cost); one analysis corner; hold damage did not appear on ASAP7 at these margins.} \\
\bottomrule
\end{tabular}
\end{table}

This evidence supports the contract and the decomposition of Section~\ref{sec:contract}, is reproducible with public tools and carries no proprietary data. It does not test the learned policy; its role is to motivate the two quantities the policy must learn: how much damage an edit creates, and which repair clears the most damage per unit of give-back.

\section{Planned Public Evaluation}\label{sec:eval}
\subsection{Reproducible experiment design}
The study uses public RTL designs and implementation scripts with documented technology and library licences \cite{openroad,opensta,asap7}: the four designs above plus further OpenROAD-flow-scripts designs for held-out families, on ASAP7 as the primary technology and on a second public technology for the transfer experiment. It spans a range of timing pressure and both clean and baseline-violating starts. All arms share functional constraints, activity assumptions, allowed cell sets, power components and acceptance criteria; unsupported checks are disclosed, and the public flow is not described as commercial signoff.

For a repair-only comparison, all policies receive the identical recovery-only checkpoint. For an end-to-end comparison, every policy starts from the same accepted checkpoint and receives the same wall-time or analysis-call budget; since raw recovery savings may then differ, final implemented power is the primary outcome, and retention is reported with its own denominator and never used alone to rank different recovery trajectories. Table~\ref{tab:arms} lists the arms. A published learned baseline is reproduced only when its code, models and action space support a fair comparison; any mismatch in objective, technology, actions or verification fidelity is disclosed, and an unmatched comparison cannot establish state-of-the-art performance.

\begin{table}[t]
\centering\footnotesize
\caption{Planned comparisons. Every arm uses matched legality rules and independent final validation; no results are implied.}
\label{tab:arms}
\begin{tabular}{@{}p{0.47\columnwidth}p{0.47\columnwidth}@{}}
\toprule
Comparison & Purpose \\
\midrule
Existing recovery and repair flow & Reproducible conventional baseline \\
Same contract, heuristic ranking & Isolate learning from scoping and budget policy \\
Same contract, random ranking & Test whether candidate intelligence matters beyond search effort \\
Reproducible published learned sizer or RL ECO policy & Compare learning methods within compatible action spaces \\
Learned ranking without uncertainty adjustment & Measure the utility and cost of calibration \\
Learned recovery only; learned repair only & Attribute benefit to the right phase \\
Frozen model; separately reported adaptation & Test transfer and the cost of new labels \\
\bottomrule
\end{tabular}
\end{table}

\subsection{Pre-registered hypotheses}
The following hypotheses and criteria are declared before the study; the numerical thresholds are proposals to be frozen with the protocol.
\begin{itemize}\setlength{\itemsep}{1pt}
\item \textbf{H1 (efficiency).} Under the identical contract and matched analysis-call budgets, learned ranking reaches the heuristic arm's final implemented power with at least 25\% fewer analysis calls, or strictly lower final power at equal calls, with paired within-design confidence intervals excluding zero on a majority of design families.
\item \textbf{H2 (safety).} Uncertainty-adjusted ranking lowers the protected-constraint regression rate and the rollback fraction by at least one third relative to unadjusted ranking, at equal or better final power.
\item \textbf{H3 (attribution).} The full learned arm's gain over the heuristic arm exceeds that of either single-phase arm, and both single-phase arms show non-negative gain; the contract alone does not explain the benefit.
\item \textbf{H4 (transfer).} A model frozen after training on other design families retains at least half of its in-family advantage over the heuristic arm, and charged online adaptation recovers part of the remainder.
\end{itemize}
Failure to meet a criterion is reported as such; arms are not dropped after the fact.

\subsection{Measurements and analysis}
For each design we report final total, leakage and dynamic power where supported; absolute saving; retention when valid; worst and total negative setup and hold slack; new and worsened checks by view; electrical and physical violations; area; inserted buffers; and wirelength or congestion where meaningful, together with the successful-transaction fraction and the reasons for rejection. Near-zero differences follow a tolerance fixed before comparison. Runtime includes candidate generation, feature extraction, inference, all successful and failed EDA trials, model updates, physical implementation and final verification; training cost is reported separately and included in any amortized-use calculation. Timing and power calls are counted and compute resources disclosed, so a reduction in prediction time cannot conceal increased verification work. Learning metrics include power-change and damage-reduction error, ranking quality, risk calibration, fallback frequency and performance on unseen families. Outcomes are compared at design level and as paired within-design differences; correlated cell trials are not treated as independent samples, and budget sweeps report failure and rollback rates rather than omitting unsuccessful settings.

\subsection{Artifacts and evidence status}
The reproducibility package will include software and library revisions, design and constraint manifests, activity generation, candidate and feature definitions, dataset splits, training and calibration scripts, model parameters, seeds, raw trial records and scripts that recreate every table and figure, each output tagged as an incremental estimate or an implemented measurement. This manuscript supplies the method, the companion study's public evidence for the contract and the protocol; it contains no executed study of the proposed model, no trained weights and no measured comparison against the listed baselines. Those are required before any quantitative claim about AI-driven power saving, speed-up, robustness or generalization.

\section{Limitations}
The learned surrogate can misrank actions when timing paths change, libraries differ or the activity profile shifts, and ensemble disagreement need not identify every error. Deterministic verification bounds what is committed, but depends on correct models, constraints, measurement fidelity and implementation checks. The framework proves neither convergence nor optimality nor the existence of a repair within a chosen budget, and conservative acceptance can discard branches a broader search would repair. Global re-checking and physical validation have a cost that caching, incremental analysis and batching may reduce only after validation against complete checks. Baseline-relative mode preserves existing violations by design and must not be confused with a tapeout-clean implementation.

\section{Conclusion}
The proposed framework makes post-route power optimization AI-driven in a specific sense: a learned, uncertainty-aware policy decides which power-recovery edits to try, which repairs to apply, in what order and when to stop, while deterministic EDA analysis remains the authority for accepting a new implementation. Its contract, baseline-relative damage sets, a saving-tied budget and complete rollback, is supported by public-design evidence from a companion study. Its central hypothesis, that learned ordering reduces unproductive trials and raises retained saving per unit of verification, is stated with a pre-registered public protocol; the next step is to run it. Until then, the contribution is a specified methodology and an evaluation plan.

\section*{AI assistance disclosure}
Large language model assistants (OpenAI ChatGPT and Anthropic Claude) assisted with manuscript restructuring, literature discovery, mathematical consistency checks and drafting of the architecture figure. All technical content, equations and claims were authored and verified by the author(s). This writing assistance is distinct from the learned optimization policy of Section~\ref{sec:policy}, which has not been trained or evaluated for this manuscript.

\balance


\begin{thebibliography}{99}\footnotesize
\bibitem{fatemi2019} H. Fatemi, A. B. Kahng, H. Lee, J. Li, and J. Pineda de Gyvez, ``Enhancing sensitivity-based power reduction for an industry IC design context,'' \textit{Integration}, vol. 66, pp. 96--111, 2019.
\bibitem{kahng2013} A. B. Kahng, S. Kang, H. Lee, I. L. Markov, and P. Thapar, ``High-performance gate sizing with a signoff timer,'' in \textit{Proc. ICCAD}, 2013, pp. 450--457.
\bibitem{hu2012} J. Hu, A. B. Kahng, S. Kang, M.-C. Kim, and I. L. Markov, ``Sensitivity-guided metaheuristics for accurate discrete gate sizing,'' in \textit{Proc. ICCAD}, 2012, pp. 233--239.
\bibitem{flach2014} G. Flach, T. Reimann, G. Posser, M. Johann, and R. Reis, ``Effective method for simultaneous gate sizing and V$_{\mathrm{th}}$ assignment using Lagrangian relaxation,'' \textit{IEEE TCAD}, vol. 33, no. 4, pp. 546--557, 2014.
\bibitem{ozdal2011} M. M. Ozdal, S. Burns, and J. Hu, ``Gate sizing and device technology selection algorithms for high-performance industrial designs,'' in \textit{Proc. ICCAD}, 2011, pp. 724--731.
\bibitem{leegupta2013} J. Lee and P. Gupta, ``ECO cost measurement and incremental gate sizing for late process changes,'' \textit{ACM TODAES}, vol. 18, no. 1, 2013.
\bibitem{ptdocs} Synopsys, \textit{PrimeTime command reference: fix\_eco\_power, fix\_eco\_leakage, fix\_eco\_timing}, version V-2023.12-SP5, 2024.
\bibitem{tempusdocs} Cadence Design Systems, \textit{Tempus User Guide, Signoff ECO: Setup Timing Recovery After Large Leakage or Total Power Optimization}, product version 25.12, 2025.
\bibitem{zahn2008} B. Zahn, ``A utility for leakage power recovery within PrimeTime SI,'' in \textit{Synopsys Users Group Boston}, 2008.
\bibitem{zahn2010} B. E. Zahn, ``System and method for employing signoff-quality timing analysis information to reduce leakage power in an electronic circuit and electronic design automation tool incorporating the same,'' U.S. Patent Application US20100050144A1, Feb. 2010.
\bibitem{openroad} The OpenROAD Project, ``OpenROAD-flow-scripts'' and ``Gate Resizer'' documentation, \url{https://openroad.readthedocs.io}, accessed Oct. 2026.
\bibitem{opensta} J. Cherry and contributors, ``OpenSTA,'' public static timing analysis software, \url{https://github.com/The-OpenROAD-Project/OpenSTA}, accessed Oct. 2026.
\bibitem{asap7} L. T. Clark \textit{et al.}, ``ASAP7: A 7-nm FinFET predictive process design kit,'' \textit{Microelectronics Journal}, vol. 53, pp. 105--115, 2016.
\bibitem{lu2020} Y.-C. Lu, S. Nath, S. S. K. Pentapati, and S. K. Lim, ``A fast learning-driven signoff power optimization framework,'' in \textit{Proc. ICCAD}, 2020.
\bibitem{dagsizer2023} C.-K. Cheng, C. Holtz, A. B. Kahng, B. Lin, and U. Mallappa, ``DAGSizer: A directed graph convolutional network approach to discrete gate sizing of VLSI graphs,'' \textit{ACM TODAES}, vol. 28, no. 4, 2023.
\bibitem{transsizer2022} S. Nath, G. Pradipta, C. Hu, T. Yang, B. Khailany, and H. Ren, ``TransSizer: A novel transformer-based fast gate sizer,'' in \textit{Proc. ICCAD}, 2022.
\bibitem{legosize2025} Y.-C. Lu, K. Kunal, G. Pradipta, R. Liang, R. Gandikota, and H. Ren, ``LEGO-Size: LLM-enhanced GPU-optimized signoff-accurate differentiable VLSI gate sizing in advanced nodes,'' in \textit{Proc. ISPD}, 2025.
\bibitem{ye2025} Y. Ye, P. Xu, L. Ren, T. Chen, H. Yan, B. Yu, and L. Shi, ``Learning-driven physically aware large-scale circuit gate sizing,'' \textit{IEEE TCAD}, vol. 44, no. 5, pp. 1901--1914, 2025.
\bibitem{pan2026} Z. Pan, Z. Guo, Y. Du, R. Wang, and Y. Lin, ``Disentangled differentiable timing-power co-optimization with quad-gradient gate sizing,'' in \textit{Proc. DAC}, 2026.
\bibitem{insta2025} Y.-C. Lu, Z. Guo, K. Kunal, R. Liang, and H. Ren, ``INSTA: An ultra-fast, differentiable, statistical static timing analysis engine for industrial physical design applications,'' in \textit{Proc. DAC}, 2025.
\bibitem{jiang2024} W. Jiang, V. A. Chhabria, and S. S. Sapatnekar, ``IR-aware ECO timing optimization using reinforcement learning,'' in \textit{Proc. MLCAD}, 2024.
\bibitem{xgboost} T. Chen and C. Guestrin, ``XGBoost: A scalable tree boosting system,'' in \textit{Proc. KDD}, 2016, pp. 785--794.
\bibitem{companion} Anonymous, ``Damage-aimed timing repair after post-route power recovery,'' companion manuscript under review, 2026.
\end{thebibliography}
\end{document}